\documentclass[letterpaper,12pt]{article}   
\usepackage{osajnl2} 
\usepackage[draft]{hyperref} 

\usepackage{graphicx}
\usepackage{url}

\usepackage{amssymb,amsmath}

\begin{document}

\title{Assessment of some experimental and image analysis factors for background oriented schlieren measurements\footnote{This paper was published in \emph{Applied Optics} and is made available as an electronic reprint with the permission of OSA. The paper can be found at the following URL on the OSA website: \texttt{http://www.opticsinfobase.org/ao/abstract.cfm?uri=ao-51-31-7554}. Systematic or multiple reproduction or distribution to multiple locations via electronic or other means is prohibited and is subject to penalties under law.}}

\author{Ardian B. Gojani$^{*}$ and Shigeru Obayashi}
\address{Institute of Fluid Science, Tohoku University, \\2-1-1 Katahira, Aoba, Sendai 980-8577, Japan}
\address{$^*$ Corresponding author: gojani@edge.ifs.tohoku.ac.jp}

\begin{abstract}Quantitative measurements of fluid flow properties can be achieved by background oriented schlieren. In this paper it is shown that this depends on several factors. Image quality index is used to investigate the influence of the image sensor and the quality of its output. Image evaluation is applied to synthetic images which are treated with a step function, so that they simulate the sharp density jump. The gradual change of the evaluated vector shift revealed the major dependence on the interrogation window, and less so on the background features. BOS applied to shock wave reflection from a wedge in a shock tube gave qualitative results, due to large uncertainties. But, the application to cooling by natural convection gave satisfactory results, comparable to thermocouple data and theory.
\end{abstract}


\maketitle

\section{Introduction}

Measurements of fluid flow properties can be achieved by numerous visualization techniques, differing in - among others, - complexity, accuracy, precission, cost, and measured property. Although flow visualization is an old and a well established method, there are always searches for advances and improvements. An oportunity is sought in simplifying quantitative measurements using schlieren based methods. As Settles testifies \cite{set}, in principle, it is possible to obtain quantitative schlieren measurements, but this is done rarely due to complexities and cost, or due to the development of interferometry, which is superior with regards to accuracy.

Recent enormous advances in digital image processing and image sensors have opened new possibilities for improving certain flow visualization techniques. A case in point is background oriented schlieren (BOS) as outlined by Meier \cite{mei} and Richard and Raffel \cite{rr}. Conceptually, BOS is a simple technique: the field of interest is photographed with and without the flow, giving the measurement and reference images, respectively. The comparison of both images, which differ due to the change in the density, consequently refractive index of the medium, yields the density gradient of the fluid in flow. This comparison can be done in numerous ways, but BOS usually employs the cross-correlation method that initially has been developed and applied for laser speckle photography and particle image velocimetry (PIV).

BOS is a combination of two techniques: image recording is done similarly to variations of schlieren that do not include a knife edge (schlieren through colored filters, grid based schlieren, etc.), while image evaluation is done in the same manner as for speckle photography or PIV. Thus, BOS lends naturally to comparisons to these techniques. Such comparisons have already been made, in particular between BOS and color schlieren: Elsinga et al. \cite{els} have studied the differences between BOS and color schlieren based on a segmented color filter, while Hargather and Settles \cite{hs} have compared BOS to color schlieren based on a continuously varying colored filter, and to standard grayscale schlieren. The outcome of these studies was that BOS might have a limited spatial resolution due to object blur, but its sensitivity is superior. A great advantage of BOS in comparison to other techniques is that it can be applied with relative ease to large scale outdoor flows, where the landscape can serve as the background, as reported by Hargather and Settles \cite{hs2}. In these cases, though, the experimenter does not have any control over the background. If, on the other hand, the experiments are conducted in a laboratory, the design of the background can be adjusted for optimal conditions. This includes determination of background features, their size and density, as well as the response in image distortion due to large density gradients from the flow. All these effects have been studied to various degree of sophistication. For example, Sj\"{o}dahl in \cite{sjo} has thoroughly analyzed the effect of the interrogation window on the cross-correlation algorithm applied to speckle photography. The standard reference on PIV, \cite{piv}, contains a chapter dedicated to image evaluation methods for PIV, which can be applied to BOS with minor modifications. These modifications should take into account the fact that in speckle photography and in PIV the imaged features are random and change between measurement and reference image. In BOS, these features, i. e. the dots, their size, density and frequency, should be conserved and any change should be attributed to the flow itself. 

In order to quantify the measurement obtained from BOS, one needs to consider several factors prior to the experiment and in preparation for image analysis. This paper will consider a few of these, namely (i) the image quality, (ii) the effect of the size of interrogation windows, and (iii) the effect of features on the imaged background. The image quality of three types of cameras is quantified using the universal image quality index, and it is tested by obtaining BOS data from two types of flows: shock wave reflection from a wedge in a shock tube, and natural convection described by the Newton's law of cooling. Image evaluation is investigated by cross-correlation of synthetically generated images subjected to the step function. The carried procedure in assessing BOS focuses only on the initial preparations for applying the technique. Once an experimenter has adopted a setup, further improvements can be achieved, and many suggestions can be found in the literature. An example is the application of colored background, which allows for eight correlations from a single background \cite{leo}.

\section{Image quality}

The speed with which a fluid flows determines the shutter speed (exposure time) that the camera needs to operate with, and this proves to be one of the most demanding factors in choosing an image sensor. Shock waves in a shock tube, for example, can be imaged only with scientific grade high-speed cameras, such as Imacon DSR200 or Shimadzu HPV-1, which are capable of imaging at times shorter than 1 $\mu$s. Slow flows that can reach a pseudosteady state, such as slow cooling by natural convection, can be imaged with standard DSLR cameras. In these later cases exposure time is not a limiting factor, because it can be large enough to reach an average value of temperature reading smaller than the measurement uncertainty, but still be orders of magnitude smaller than the temperature measurement steps. Thus, one would be free to choose a camera with a high pixel count or a large sensor size, so that the image detail is satisfactory for precise measurements.

\begin{figure}
\includegraphics{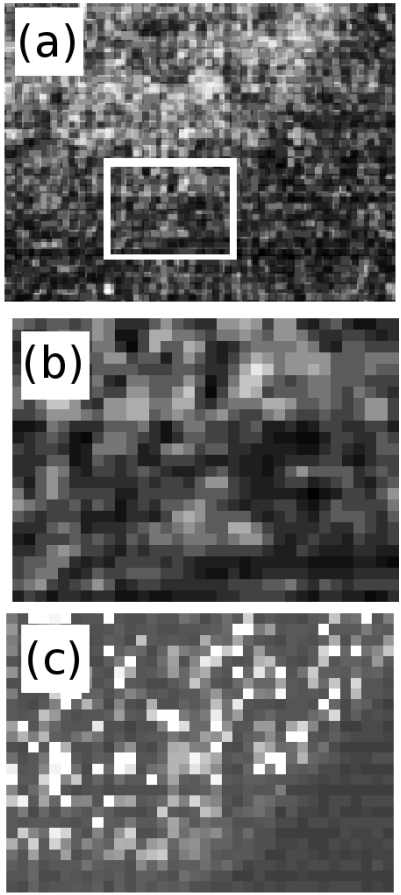}
\caption{Images (a) and (b) - the latter being the framed part in the former, - are taken with Imacon DSR200 camera, and (c) with Shimadzu HPV-1. Images (a) and (c) show the same field of view, while (b) and (c) have the same pixel count. This detail of the entire image corresponds to the small rectangle starting at the fiftieth pixel in Fig. \ref{ccshock}.} \label{pixels}
\end{figure}

\begin{figure}
\includegraphics{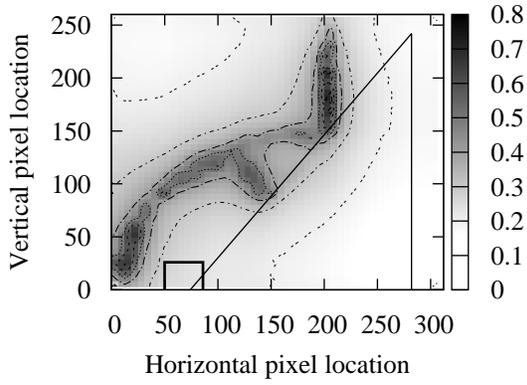}
\caption{Background oriented schlieren result for the shock reflection from a wedge (black full lines), shown as a magnitude map of vector shift, with pixel values in coordinates and pixel shift amount in the magnitude bar. Contours border zones with 0.1 pixel shift difference.} \label{ccshock}
\end{figure}

Images shown in Fig. \ref{pixels} are from both types of high speed cameras and show a detail of the same background imaged through the test section of a shock tube. If one considers the sensitivity of a BOS setup based on the geometry of the layout and the size of the image sensor, Imacon camera would be preferable to Shimadzu, for two reasons: larger pixel count ($1200 \times 980$ pixels for Imacon vs. $312 \times 260$ pixels for Shimadzu), and smaller pixel size ($\approx 10 \mu$m vs. $\approx 60 \mu$m). But, despite the fact that the shown field of view is the same and that the images were taken with the same lens as well as under the same illumination, obtained images are quite different, which fundamentally comes about due to the different quantum efficiencies of the respective image sensors. A dramatic outcome of this difference is that the direct (without any processing) evaluation with cross-correlation of images captured by Imacon camera could not give any meaningful results, while images captured by Shimadzu did, as shown in Fig. \ref{ccshock}. 

\begin{figure}
\includegraphics{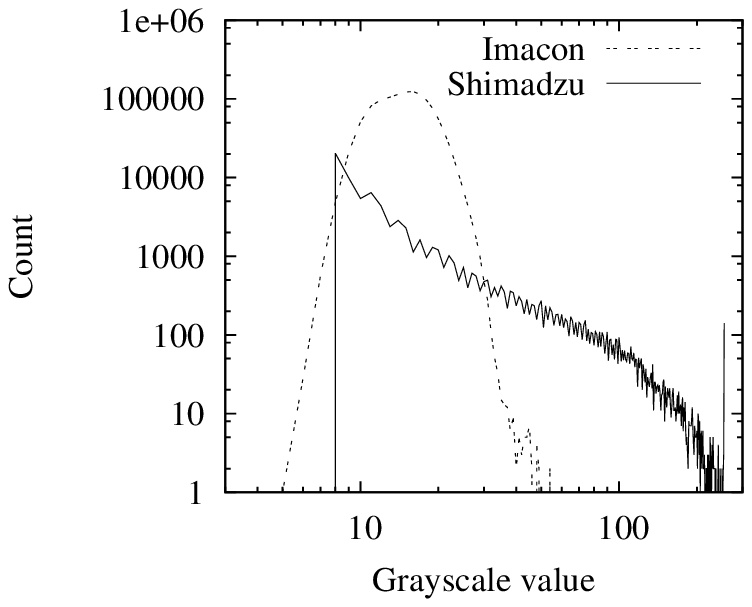}
\caption{Histogram of the Imacon and Shimadzu camera, operated under the same conditions.} \label{hist}
\end{figure}

The difference can be explained through different response to luminance of the image sensors and the contrast values of the output file. Both cameras have a sensor with 10 bit dynamic range and give comparable dark images, but the histogram of the measurement images, as given in Fig. \ref{hist}, shows that the Shimadzu camera produces a better contrast. Calculation of the universal image quality index, as proposed by Wang and Bovik \cite{wb}, reveals that not only contrast, but also the response to luminance plays a role. In determining the image quality index, an image is considered as a vector of grayscale values $x_i$, for which one can calculate the average 
\begin{equation*}
\bar{x}=\frac{1}{N}\sum_{i=1}^N x_i
\end{equation*}
and the standard deviation
\begin{equation*}
\sigma_x=\frac{1}{\sqrt{N-1}}\sum_{i=1}^N \left(x-\bar{x}\right),
\end{equation*}
where $N$ is the pixel count. The original universal image quality index is defined as the product of three terms: one describing the correlation between compared images, the other descibing the response to luminance, and the final term describing the contrast. Since the backgrounds used in BOS are high frequency random distribution of dots, the term that describes the correlation is always very close to zero and does not play a role in determining image quality. Hence, we introduce a modified universal image quality index, which takes into account only the comparison of luminance and contrast between two images, and is defined as 
\begin{equation} \label{Q}
Q_{mod}=Q_L \cdot Q_C
\end{equation}
with
\begin{equation}
\begin{split}
Q_L &=\frac{2\bar{x}\bar{y}}{\bar{x}^2+\bar{y}^2},\\
Q_C &=\frac{2\sigma_x \sigma_y}{\sigma_x^2+\sigma_y^2},
\end{split}
\end{equation}
where $x$ is the image for which the quality index is being determined, and $y$ is a reference image. In our evaluations of $Q_{mod}$, $y$ is the digitally produced background. The first term of Eq. \ref{Q} is related to the luminance of the images $x$ and $y$, and measures how close are average luminance values between these images ($Q_L=0$ for very different luminance values, $Q_L=1$ for the same luminance values). The second term depends on the standard deviation of the images, thus it describes how similar are the contrast values ($Q_C=0$ for very different contrast values, $Q_C=1$ for similar contrast values of the compared images).

\begin{table} 
\centering
\begin{tabular}{r c c c c}
\hline
& $Q_L$ & $Q_C$ & $Q_{mod}$\\
\hline
Imacon DSR200 & 0.18 & 0.07 & 0.01\\
Shimadzu HPV-1 & 0.62 & 0.54 & 0.33\\
Pentax K-5 & 0.99 & 0.71 & 0.70\\
\hline
\end{tabular}
\caption{Components of the image quality index describing luminance $Q_L$ and contrast $Q_C$ for three cameras used in experiments.}
\label{q-values}
\end{table}

The values for the modified universal image quality index $Q_{mod}$, as well as its individual terms $Q_L$ and $Q_C$ for tested cameras (Imacon DSR200, Shimadzu HPV-1, and Pentax K-5) are given in Table \ref{q-values}. The imaging of shock wave reflection by a wedge in a shock tube is imaged by high speed cameras that yield images of modest quality, as can be judged by the values of $Q_{mod}$. On the other hand, the imaging of natural convection was done with a Pentax K-5 DSRL camera, and, as it is expected, it performs much better, demonstrated by the high values of the $Q_L$ and $Q_C$  and their overall product. Therefore, a preliminary investigation of the cameras used for BOS can be done by determining $Q_{mod}$: a fixed value that would qualify an image as useful or not-useful for image analysis is impossible to be given, but a reasonable judgement can be given based on how close $Q_L$ and $Q_C$ are to 1, and the main factor effecting low image quality (luminance or contrast) can be diagnosed. 

\section{Interrogation window size}

\begin{figure}
\includegraphics{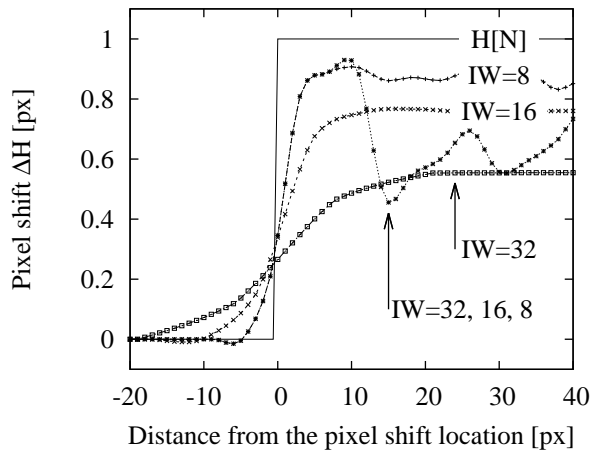}
\caption{The resolution of the evaluation of a one pixel shift as a function of different sizes of IW. The case for IW=32, 16, 8 pixels shows the multipass evaluation.} \label{srf}
\end{figure}

Image evaluation by cross-correlation consists on defining subsets of the measurement image, representing the interrogation window (IW), and comparing their intensity fields to all equal in size subsets in the reference image. The output of cross-correlation for a prearranged IW is a vector, with magnitude and direction corresponding to the shift of the correlation peak. The number of independent vectors, thus, the evaluated spatial resolution, depends on the size of the IW. This is ilustrated for the step function $H[N]$ in Fig. \ref{srf}, where 
\begin{equation}
H[N]= \left\{ 
   \begin{array}{l l}
     0 & \quad N < 0, \\
     1 & \quad N \geq 0.\\
   \end{array} \right.
\end{equation}
Similar to the work of K\"{a}hler et al. \cite{kah}, half of a $128 \times 128$ pixels synthetically generated reference image of randomly distributed dots is shifted for a pixel, giving the simulated measurement image. The cross-correlated evaluation was done using the PIV plugin for ImageJ \cite{q}, \cite{ij}. Image evaluation with different IW shows that the smaller the IW the closer the evaluated jump is to the real step function. The spread and gradual increase of the discontinuity means that there can not be an independent shift vector within the length it takes the evaluated pixel shift value to jump. Therefore, the evaluated spatial resolution for IW of side length equal to 32 pixels is about 38 pixels, for IW=16 is 30, and for IW=8 it is about 16 px. This explains the \textit{spill} of pixel shifts inside the wedge, shown in Fig. \ref{ccshock}. The IW applied to the images from the shock tube experiment were of 8 px, and the resolution is estimated to be at least 10 px. Since the optical magnification for this setup was 0.1, the measurement uncertainty then is 5 mm.

\begin{figure}
\includegraphics{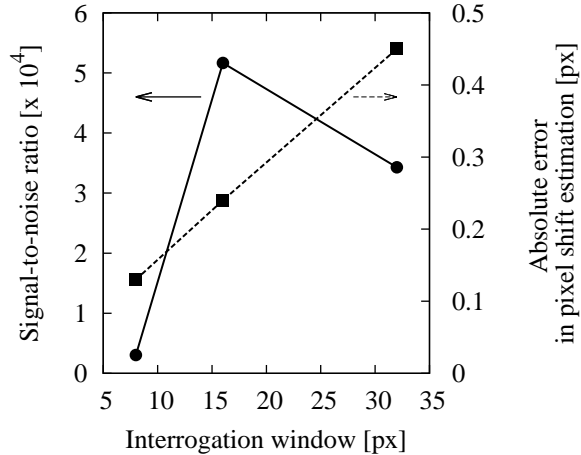}
\caption{Signal-to-noise ratio and absolute error in pixel shift estimation calculated for the cases in Fig. \ref{srf}.} \label{snr}
\end{figure}

Calculations shown in Fig. \ref{srf} reveal also the effect of the IW on the signal-to-noise ratio (SNR), as well as the absolute error in estimaging pixel shift, a result which is shown in Fig. \ref{snr}. SNR is calculated as the inverse of the square of the standard deviation, \cite{hs2}, while the error in estimation of the pixel shift is the difference from 1 of the mean pixel shift value obtained from cross-correlation after the shift location. Since these calculations are not done for optimal match between background features and IW, Fig. \ref{snr} is shown only for illustrative purposes: one can find a maximal value for SNR, while the error in pixel shift always increases with the increase of IW.

\section{Background features}

\begin{figure}
\includegraphics{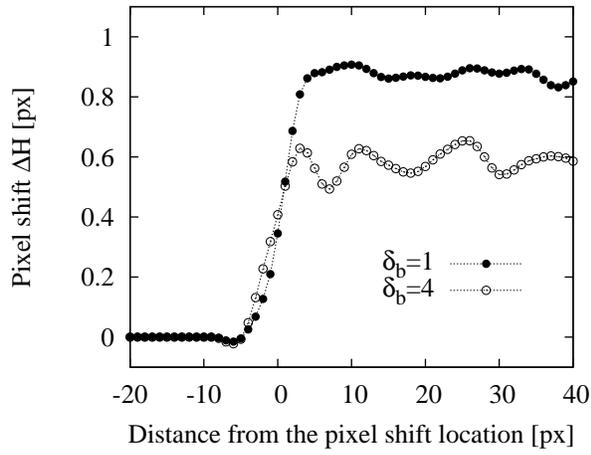}
\caption{The effect of dot size on evaluating pixel shift.} \label{size}
\end{figure}

The preparation of the background involves determination of the dot size, frequency (in terms of the number of dots in the field of view), and coverage. For a want of a better analogy, one can say that the dots in the background serve the same role as the ticks in a ruler. Fig. \ref{size} shows the effect of the dot size in the resolution, where dot sizes $\delta_b$ equal to one and four pixels are compared after images are treated by a one pixel step function. The evaluation does not show any large effect on the resulution, but it does give different values for the pixel shift. For the case with $\delta_b=4$ px, the effective shift of the image corresponds to a quarter of a dot, hence some pixels are not recorded as shifted. This situation appears in synthetic image evaluations, because these images have a well defined binary structure. In experiments, an image of a binary background results in a grayscale image with spread histogram peaks around the binary values. This leads to more accurate results during evaluation.

\begin{figure}
\includegraphics{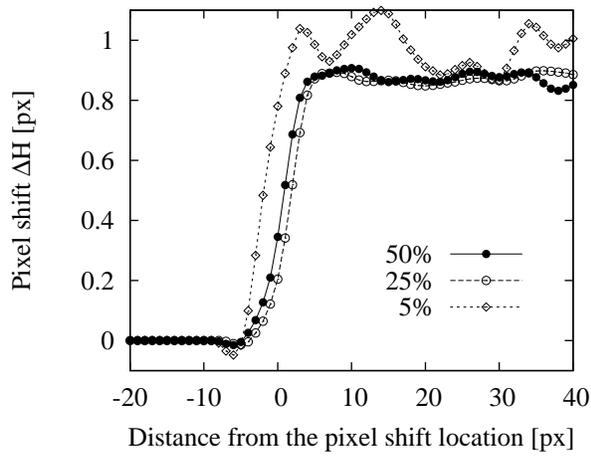}
\caption{The effect of dot distribution density on evaluating pixel shift.} \label{density}
\end{figure}

Fig. \ref{density} shows the effect of dot density, which is defined as the number of dots per unit area of the field of view projected in the total image area. Maximal dot density, 50\%, means that half of the image is covered by dots, and the minimal density simulated (5\% coverage with dots) is mainly a white featureless background. As previously, the image is treated to a one pixel shift and evaluated with IW=8 px. Virtually, there is no difference in evaluating images with dot density of higher than 20-25\%, but pixel shift evaluation artifacts start showing up for images with lower number of dots. This result is in agreement with the previously published requirement that an IW should have at least four to five dots, each covering 2 pixels \cite{piv}.

\section{Quantifying natural convection}

\begin{figure}
\includegraphics{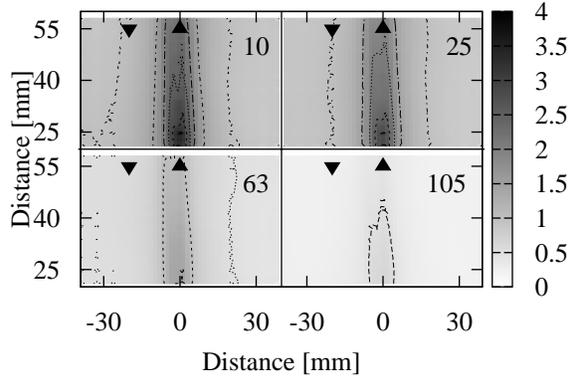}
\caption{BOS results for cooling by natural convection, taken at four different times: 10 s, 25 s, 63 s, and 105 s after the heater was turned off (times noted in the upper right courner of each map). Thermocouple positions are noted by the symbols $\blacktriangledown$ and $\blacktriangle$. Coordinates give the distance from the center of the heat source, in mm, and the magnitude bar gives the pixel shift. Contour lines correspond to borders between zones with pixel shift diference of 0.5 pixel.} \label{nc4}
\end{figure}

\begin{figure}
\includegraphics{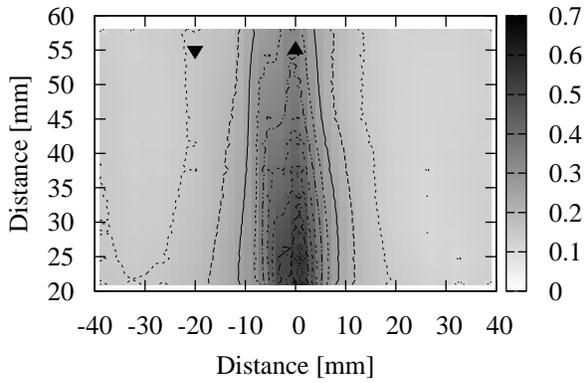}
\caption{A detailed image taken at 105 s after the heater was turned off (BOS result shown in the right lower corner of Fig. \ref{nc4}). In this case, contour lines border zones with 0.05 pixel shift difference.} \label{nc1}
\end{figure}

While the results for density gradient measurements behind a shock wave in a shock tube, as shown in Fig. \ref{ccshock}, are very coarse, mainly due to the low pixel count of the high speed camera, the results for BOS applied to natural convection show fine detail, as given in Figs. \ref{nc4} and \ref{nc1}. A 100 W horizontally placed radiating cylinder was enclosed in a box, and at 55 mm above it, along 60 mm in a plane parallel to the heater, were placed 2 thermocouples. Figure \ref{nc4} shows the evolution of the temperature field with time as a map of pixel shift values for four different times: 10 s, 25 s, 63 s, and 105 s after the heater was turned off. Contour lines correspond to zones with pixel shift difference of 0.5 pixel. A more detailed view for the temperature field at 105 s after the heat source was turned off is given in Fig. \ref{nc1}, in which case zones with pixel shift difference of 0.05 pixel are outlined. In both figures, the position of two thermocouples is designated by symbols $\blacktriangledown$ and $\blacktriangle$. The zone around the radiating cylinder can not be shown, because the holder of the cylinder obscures the view.

\begin{figure}
\includegraphics{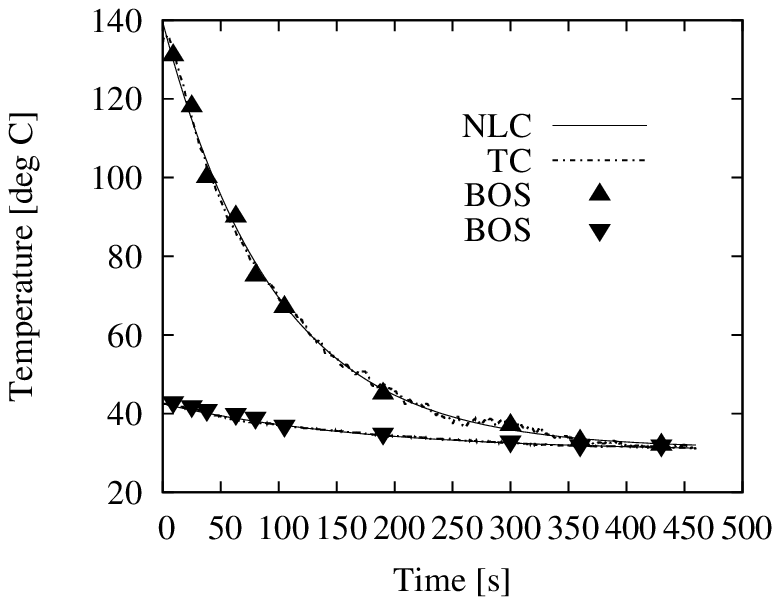}
\caption{Quantitative BOS results for cooling by natural convection, and the comparison of BOS data to thermocouple (TC) readings and Newton's law of cooling (NLC).} \label{tc}
\end{figure}

The comparison of temperature evolution by BOS, thermocouple readings, and according to Newton's law of cooling shows a satisfactory agreement, as shown for both thermocouples in Fig. \ref{tc}. BOS imaging was done with an extra large pixel count camera (16 megapixels), which had several benefits: a dot covered $8 \times 8$ px, the dots had five distinct grayscale values, and IW was $16 \times 16$ px. It should be noted that for the sake of computational speed, the image was reduced in size 4 times per direction, giving two dots per pixel. The dot density was about 50\%, with about 10\% of dots concentrating around each grayscale value (0, 43, 86, 128). The magnification achieved by the optical setup was 0.15, with 30 pixels covering 1 mm of the field of view. The limit on the spatial resolution based on the IW (in this case, 30 px) and the pixel shift equal to 0.01 px in the neighbourhood of the thermocouples are of the same value, namely 1 mm, defining the overall spatial resolution of the measurement.

Applying the ideal equation of state for air, as well as the Gladstone-Dale relation for the functional relationship between the refractive index and the density of air, it was obtained that 1 pixel shift corresponds to 60 $^{\circ}$C temperature change. Since cross-correlation in BOS can determine pixel shifts with accuracy of 0.01 pixel, one can conclude that the measurenment sensitivity of BOS can be considered to be of the same order of magnitude to that of the thermocouples (0.1 $^{\circ}$C).

\section{Conclusion}

In this paper is given an assessment of background oriented schlieren from the image recording and image analysis point of view. Also, a summary of results of BOS applied to two very different flow fields: shock wave reflection and natural convection, is given. Imaging of these flows required different equipment, thus the obtained results differed in quality. This paper proposed the modification and utilization of the universal image quality index $Q_{mod}$ as an initial and simple step for assessing a BOS system. Although a fixed value of the index that would define a useful image is impossible to be given, an experimenter can make reasonable judgements based on the value of $Q_{mod}=Q_L Q_C$ whether a BOS setup would yield images appropriate for evaluation by cross-correlation. Since shock wave phenomena require high speed imaging, the cameras used for the experiment could reach limited resolution. As a consequence, only qualitative results were possible. Cooling by natural convection, in turn, was measured by a camera that produced images with high quality index, hence quantitative results were obtained. 

In addition, this paper presented some simple results of BOS image analysis. Quantification of BOS images is mainly done with cross-correlation, and image evaluation based on that procedure is characterized by two conflicting requirements:
\begin{itemize}
\item for accurate results, the interrogation window for cross-correlation should be as small as possible, so that averaging does not cause loss of information; but,
\item for good signal-to-noise ratios, interrogation window should be as large as possible, so that enough detail is included in the evaluation.
\end{itemize} 
As a result, a smaller interrogation window gave lower uncertainty in the measurement of position (higher spatial resolution), but still it was impossible to obtain independent vectors of a pixel shifted image with higher frequency than a vector per 8-10 pixels. Multipass (iterative) processing was also tested, but it did not show any real benefit. Since in all our experiments the interrogation window included several dots, the signal-to-noise ratio was not a determining factor for choosing an appropriate interrogation window, but it was rather the spatial resolution which affected the uncertainty of results. 

In accordance to previous studies, the presented assessment of background features confirms that a good cross-correlation evaluation requires at least four to five dots for an interrogation window, with a dot covering at least two pixels. The evaluation of the images with low dot density showed artifacts, more so in spatial resolution and less in pixel shift evaluation. It is estimated that a good background is one with at least 20\% coverage by dots. Dot size did not show any difference in terms of spatial resolution, but it gave an incorrect value for pixel shift, which is due to the effective pixel shift.

\section*{Acknowledgments}
This work was supported by Global COE Program "World Center of Education and Research for Trans-disciplinary Flow Dynamics," Tohoku University, Japan.

\end{document}